\documentclass[journal]{IEEEtran}
\usepackage{amsmath,amssymb,amsthm}
\usepackage{graphicx}
\usepackage[caption=false,font=footnotesize]{subfig}
\usepackage{algorithm}
\usepackage{algpseudocode}
\usepackage{booktabs}
\usepackage{placeins}
\usepackage{cite}
\theoremstyle{plain}
\newtheorem{proposition}{Proposition}
\theoremstyle{definition}
\newtheorem{remark}{Remark}

\usepackage[colorlinks]{hyperref}

\begin{document}
\title{\LARGE Two-Timescale Channel Estimation and Port Selection for Fluid Antenna Systems With Limited Port Observations}
\author{Xusheng~Zhu, Kai-Kit~Wong,~\IEEEmembership{Fellow,~IEEE}, Jiajian Wang, Songling Zhang, Shihao Fan%
\thanks{(\emph{Corresponding author: Kai-Kit Wong}.)}%
\thanks{X. Zhu, K. K. Wong, and J. Wang are with the Department of Electronic and Electrical Engineering, University College London, London WC1E 6BT, U.K. K. K. Wong is also affiliated with the Department of Electronic Engineering, Kyung Hee University, Yongin-si, Gyeonggi-do 17104, Republic of Korea (e-mail: \{xusheng.zhu; kai-kit.wong; jiajian.wang.25\}@ucl.ac.uk).}
\thanks{S. Zhang is with the College of Information Science and Electronic Engineering, Zhejiang University, Hangzhou, China (e-mail: sl-zhang@zju.edu.cn).}
\thanks{S. Fan is with the School of Automation and
Intelligence, Beijing Jiaotong University, Beijing 100044, China (e-mail:
fsh20028@gmail.com).}

}
\maketitle

\begin{abstract}
Port selection in fluid antenna systems (FAS) requires instantaneous channel state information, but a single radio-frequency chain can probe only a few ports in each coherence block. Existing FAS channel estimators, even with learned or generative priors, work within a single block and are limited by its few probes, which cannot resolve several propagation paths. To overcome this limitation, we propose a two-timescale channel estimation and port selection scheme that exploits the slow variation of the path angles. Specifically, a sequential Monte Carlo filter accumulates the angle posterior over blocks, while the fast-fading path gains are integrated out in closed form. Furthermore, a Fisher information analysis reveals that a single probe per block carries no angle information, whereas two or more spread-out probes accumulate it linearly over blocks. Simulation results demonstrate that the proposed scheme approaches the selection with known angles after about twenty blocks and, with four probes per block, matches memoryless schemes with eight.
\end{abstract}


\section{Introduction}
\IEEEPARstart{F}{luid} antenna systems (FAS) obtain spatial diversity from a single radio-frequency (RF) chain by switching a radiating element among many densely spaced ports of a given aperture \cite{wong2021fas}. The resulting gains in outage and rate have been characterized in \cite{zhu2026geometric,zhu2026ofdm}. However, these gains rely on the selection of a good port, which requires knowledge of the channels at all ports.

The acquisition of the required channel knowledge is a major bottleneck of FAS. As a receiver with one RF chain observes only one port per pilot symbol, the training of all ports would consume most of the coherence block. In practice, only a few ports are probed, and the channels of the remaining ports are reconstructed from these probes. To this end, existing estimators exploit different forms of prior knowledge of the channel. Specifically, linear minimum mean-square error (LMMSE) estimation relies on the spatial correlation between ports \cite{skouroumounis2023lmmse}, while sparse reconstruction exploits the multipath structure \cite{xu2024mmwave}. Alternatively, learning-based methods predict the best port from a few observed ports \cite{chai2022port}, and diffusion models serve as generative channel priors \cite{tang2025diffusion}.

\begin{figure}[t]
\centering
\includegraphics[width=\columnwidth]{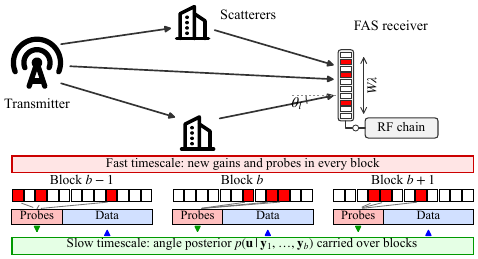}
\caption{System model and probing protocol. The probed ports of each block are shown in red.}
\label{fig:sys}
\end{figure}
Nevertheless, each of these estimators works within a single coherence block. As a result, the estimation accuracy is limited by the few probes of one block, which cannot resolve the path angles that determine the deep fades of the unprobed ports. To exploit past blocks, the channel has been predicted from its temporal correlation \cite{zhang2023fast}. However, the temporal correlation of the channel vanishes once the fading is independent across blocks, and the prediction provides no information. Even under independent fading, the path angles persist, since they change much more slowly than the small-scale fading \cite{heath2016mmwave}. Along this line, the two-timescale structure of the channel has been exploited for massive multiple-input multiple-output systems in \cite{xie2017sbem}, where all antennas are observed in every block. In contrast, a FAS receiver observes only a few ports per block, and the new path gains of every block randomize the phase of each probe. Accordingly, it remains unclear whether and how fast the path angles can be learned from these probes across blocks.

To address this issue, we develop a two-timescale channel estimation and port selection scheme, whose angle memory accumulates the angle information of all past blocks to predict the unprobed ports of every new block. The main contributions are summarized as follows. 1) We propose a sequential Monte Carlo filter that accumulates the angle posterior over blocks, where the fast-fading path gains are integrated out in closed form. Each block selects the port with the smallest posterior outage probability. 2) We derive the Fisher information of the angle per block in closed form for a single path. The resulting expression reveals that one probe per block carries no angle information, whereas two or more spread-out probes accumulate information linearly over blocks. 3) Simulation results validate the analysis. The proposed scheme with four probes per block achieves an outage probability comparable to that of memoryless schemes with eight probes. Moreover, the gain persists under an unknown number of paths, clustered scattering, and drifting angles.

\section{System Model}
\subsection{Signal Model}
As shown in Fig.~\ref{fig:sys}, a single-antenna transmitter sends unit-power pilot symbols followed by data symbols in every coherence block. At the receiver, a linear FAS of $N$ ports is uniformly spaced over $W$ wavelengths, and all ports share one RF chain. As a consequence, only one port can be observed per pilot symbol. For $L$ propagation paths between the transmitter and the FAS, the channel vector of all ports in block $b$ is given by
\begin{equation}\label{eq:channel_h}
\mathbf{h}_b=\sum_{l=1}^{L} g_{b,l}\,\mathbf{a}(u_l)=\mathbf{A}(\mathbf{u})\,\mathbf{g}_b,
\end{equation}
where $\mathbf{g}_b\sim\mathcal{CN}(\mathbf{0},\mathbf{I}_L/L)$ collects the circularly symmetric complex Gaussian path gains of block $b$, and the $n$\nobreakdash-th entry $h_{b,n}$ of $\mathbf{h}_b$ is the channel of port $n$. The path gains are independent across blocks and model the fast small-scale fading. For simplicity, all paths have equal power, and every port has unit average power. Moreover, the response matrix of the $L$ paths and the port response to one path are defined as
\begin{subequations}\label{eq:channel}
\begin{align}
\mathbf{A}(\mathbf{u})&=[\mathbf{a}(u_1),\ldots,\mathbf{a}(u_L)],\label{eq:channel_A}\\
\mathbf{a}(u)&=\big[e^{j2\pi x_1 u},\ldots,e^{j2\pi x_N u}\big]^{T},\label{eq:channel_a}
\end{align}
\end{subequations}
where $\mathbf{u}=[u_1,\ldots,u_L]^{T}$ collects the directional cosines $u_l=\sin\theta_l$ of the paths with the arrival angles $\theta_l$, $j$ is the imaginary unit, and $x_n=(n-1)W/(N-1)$ is the position of the $n$\nobreakdash-th port normalized by the wavelength.

Unlike the path gains, the angles stay constant over many blocks, as they are determined by the positions of the transmitter and the scatterers \cite{heath2016mmwave}. Therefore, the two factors in (\ref{eq:channel_h}) vary on different timescales in Fig.~\ref{fig:sys}. We assume that every directional cosine is uniformly distributed over $[-1,1]$ to model isotropic three-dimensional scattering.

Each block starts with the probing of $k$ ports in the set $\mathcal{S}_b$, which is drawn uniformly at random and independently of the received signals. Based on (\ref{eq:channel_h}), the received pilot vector can be expressed as
\begin{equation}\label{eq:pilot}
\mathbf{y}_b=\boldsymbol{\Phi}_b\,\mathbf{g}_b+\mathbf{w}_b,
\end{equation}
where $\boldsymbol{\Phi}_b=[\mathbf{A}(\mathbf{u})]_{\mathcal{S}_b,:}$ and $\mathbf{w}_b\sim\mathcal{CN}(\mathbf{0},\sigma^2\mathbf{I}_k)$ denote the rows of $\mathbf{A}(\mathbf{u})$ indexed by $\mathcal{S}_b$ and the noise, respectively, where $1/\sigma^2$ is the pilot signal-to-noise ratio (SNR).

\subsection{Port Selection and Outage}
After the probing, the receiver selects one port $\hat n_b$ for data reception, and an outage occurs if
\begin{equation}\label{eq:outage}
|h_{b,\hat n_b}|^2<\gamma,
\end{equation}
where $\gamma=(2^{R_0}-1)/\rho$ for the target rate $R_0$ and the average data SNR $\rho$. The outage-minimizing selection can be written as
\begin{equation}\label{eq:optsel}
\hat n_b=\arg\max_{n}\ \Pr\big(|h_{b,n}|^2\ge\gamma\,\big|\,\mathbf{y}_1,\ldots,\mathbf{y}_b\big),
\end{equation}
where the probability is conditioned on all pilot vectors up to block $b$, since the past blocks share the angles of the current channel.

\section{Proposed Two-Timescale Scheme}
\subsection{Slow-Timescale Angle Posterior}
Without knowledge of the angles and the number of paths, the receiver adopts an over-parameterized model with $\hat L\ge L$ paths.\footnote{The model contains the true channel if $\hat L$ is a multiple of $L$, since $\hat L/L$ coinciding model paths of power $1/\hat L$ carry the power $1/L$ of one true path.} For given model angles, the pilot vector (\ref{eq:pilot}) is zero-mean Gaussian as
\begin{equation}\label{eq:ymodel}
\mathbf{y}_b\,|\,\mathbf{u}\sim\mathcal{CN}\big(\mathbf{0},\boldsymbol{\Phi}_b\boldsymbol{\Phi}_b^{H}/\hat L+\sigma^2\mathbf{I}_k\big),
\end{equation}
where $\mathbf{u}\in[-1,1]^{\hat L}$ and $\boldsymbol{\Phi}_b$ refer to the model, and the model gains follow $\mathcal{CN}(\mathbf{0},\mathbf{I}_{\hat L}/\hat L)$ to keep unit power per port. The corresponding log-density is written as
\begin{equation}\label{eq:logdens}
\log p(\mathbf{y}_b\,|\,\mathbf{u})=-k\log\pi-\log\det\mathbf{C}_b-\mathbf{y}_b^{H}\mathbf{C}_b^{-1}\mathbf{y}_b,
\end{equation}
where $\mathbf{C}_b=\boldsymbol{\Phi}_b\boldsymbol{\Phi}_b^{H}/\hat L+\sigma^2\mathbf{I}_k$ is the covariance in (\ref{eq:ymodel}). By the matrix determinant lemma and the matrix inversion lemma, the determinant, the inverse, and the quadratic form in (\ref{eq:logdens}) become
\begin{subequations}\label{eq:woodbury}
\begin{align}
\det\mathbf{C}_b&=\sigma^{2k}\det\Big(\mathbf{I}_{\hat L}+\frac{\boldsymbol{\Phi}_b^{H}\boldsymbol{\Phi}_b}{\hat L\sigma^2}\Big),\label{eq:woodbury_det}\\
\mathbf{C}_b^{-1}&=\frac{1}{\sigma^2}\Big(\mathbf{I}_k-\boldsymbol{\Phi}_b\mathbf{M}_b^{-1}\boldsymbol{\Phi}_b^{H}\Big),\label{eq:woodbury_inv}\\
\mathbf{y}_b^{H}\mathbf{C}_b^{-1}\mathbf{y}_b&=\frac{\|\mathbf{y}_b\|^2-\mathbf{c}_b^{H}\mathbf{M}_b^{-1}\mathbf{c}_b}{\sigma^2},\label{eq:woodbury_quad}
\end{align}
\end{subequations}
where $\mathbf{M}_b=\hat L\sigma^2\mathbf{I}_{\hat L}+\boldsymbol{\Phi}_b^{H}\boldsymbol{\Phi}_b$, and $\mathbf{c}_b=\boldsymbol{\Phi}_b^{H}\mathbf{y}_b$ correlates the probes with the model responses. The constants $k\log\pi$ and $k\log\sigma^2$ cancel in the normalized posterior, and the substitution of (\ref{eq:woodbury}) into (\ref{eq:logdens}) yields the log-likelihood of block $b$ as
\begin{equation}\label{eq:loglik}
\ell_b(\mathbf{u})=\underbrace{-\log\det\!\Big(\mathbf{I}_{\hat L}+\frac{\boldsymbol{\Phi}_b^{H}\boldsymbol{\Phi}_b}{\hat L\sigma^2}\Big)}_{\text{(I)}}\underbrace{-\frac{\|\mathbf{y}_b\|^2-\mathbf{c}_b^{H}\mathbf{M}_b^{-1}\mathbf{c}_b}{\sigma^2}}_{\text{(II)}},
\end{equation}
where term (I) normalizes the density and lowers the likelihood of the angles whose responses have a low mutual coherence at the probed ports, since such angles can explain more observations. In contrast, term (II) measures the fit to the probes. Given that $\|\mathbf{y}_b\|^2$ is common to all angles, only the weighted correlation energy $\mathbf{c}_b^{H}\mathbf{M}_b^{-1}\mathbf{c}_b$ varies, and it favors the angles whose responses match the probes. As the gains are independent across blocks, the pilot vectors are conditionally independent given the angles, and the Bayes rule yields the angle posterior after $b$ blocks as
\begin{equation}\label{eq:post}
p(\mathbf{u}\,|\,\mathbf{y}_1,\ldots,\mathbf{y}_b)\propto p(\mathbf{u})\prod_{i=1}^{b}e^{\ell_i(\mathbf{u})},
\end{equation}
where $p(\mathbf{u})$ is the uniform prior. The angle posterior forms the angle memory in Fig.~\ref{fig:sys}. Each factor in (\ref{eq:post}) is large at several angles, since a few probes are consistent with more than one set of angles. As the probe sets differ across blocks, the product keeps only the angles that are consistent with all blocks. Consequently, the angle memory becomes more concentrated as blocks accumulate and removes the ambiguity of a single block.

\subsection{Fast-Timescale Port Selection}
The selection rule (\ref{eq:optsel}) requires the posterior of the current channel. Since the past probes inform the current channel only through the angles, the channel posterior is found in two steps, first for given angles and second over the angle memory (\ref{eq:post}).

\subsubsection{Port Posterior for Given Angles}
For given angles, (\ref{eq:pilot}) is a linear Gaussian model of the current gains, and the Bayesian linear estimator yields the gain posterior as
\begin{equation}\label{eq:gpost}
\mathbf{g}_b\,|\,\mathbf{y}_b,\mathbf{u}\sim\mathcal{CN}\big(\mathbf{M}_b^{-1}\mathbf{c}_b,\ \sigma^2\mathbf{M}_b^{-1}\big).
\end{equation}
Every port channel is a linear combination of the gains, and (\ref{eq:gpost}) leads to the port posterior as
\begin{equation}\label{eq:portpost}
h_{b,n}\,|\,\mathbf{y}_b,\mathbf{u}\sim\mathcal{CN}(\mu_n,v_n),
\end{equation}
where $\mu_n=\mathbf{a}_n^{T}\mathbf{M}_b^{-1}\mathbf{c}_b$ and $v_n=\sigma^2\mathbf{a}_n^{T}\mathbf{M}_b^{-1}\mathbf{a}_n^{*}$ follow from (\ref{eq:gpost}), and $\mathbf{a}_n=[e^{j2\pi x_n u_1},\ldots,e^{j2\pi x_n u_{\hat L}}]^{T}$ is the response of port $n$ to the model paths. Thus, the port amplitude is Rician with the deterministic component $|\mu_n|$ and the scattered power $v_n$, and the probability that port $n$ supports the target rate is given by
\begin{equation}\label{eq:marcum}
q_n(\mathbf{u})=Q_1\Big(\sqrt{2|\mu_n|^2/v_n},\sqrt{2\gamma/v_n}\Big),
\end{equation}
where $Q_1(\cdot,\cdot)$ is the first-order Marcum Q-function. Unlike the predicted power $|\mu_n|^2$ alone, (\ref{eq:marcum}) also accounts for the uncertainty $v_n$ and can prefer a well-predicted port.

\subsubsection{Particle Implementation of the Angle Posterior}
The second step averages (\ref{eq:marcum}) over the angle posterior (\ref{eq:post}). Since a few probes leave the angles ambiguous, the posterior has several peaks, and a grid search becomes impractical for more than one path. Instead, the posterior is approximated by $P$ particles, which are updated in every block following the resample-move approach of \cite{chopin2002smc}. First, the particles are weighted by the new likelihood (\ref{eq:loglik}) and resampled. Second, every particle is moved $T$ times to keep the particles diverse after resampling. Each move either adds small Gaussian noise to the angles or draws one angle again from the prior, and the candidate $\tilde{\mathbf{u}}^{(p)}$ replaces particle $\mathbf{u}^{(p)}$ with the probability defined as
\begin{equation}\label{eq:mh}
\alpha^{(p)}=\min\Big\{1,\exp\Big(\sum_{i=1}^{b}\big[\ell_i(\tilde{\mathbf{u}}^{(p)})-\ell_i(\mathbf{u}^{(p)})\big]\Big)\Big\},
\end{equation}
which involves the likelihoods of all past blocks and keeps the particles consistent with the whole history.

Finally, the average of (\ref{eq:marcum}) over the particles approximates the probability in (\ref{eq:optsel}), and the selection rule is formulated as
\begin{equation}\label{eq:select}
\hat n_b=\arg\max_{n}\ \frac{1}{P}\sum_{p=1}^{P}q_n\big(\mathbf{u}^{(p)}\big).
\end{equation}

\begin{algorithm}[t]
\caption{Proposed Two-Timescale Scheme}
\label{alg:smc}
\begin{algorithmic}[1]
\Require $k$, $\hat L$, $P$, $T$, $\gamma$, $\sigma^2$
\Ensure $\hat n_b$ for $b=1,2,\ldots$
\State Draw $\{\mathbf{u}^{(p)}\}_{p=1}^{P}$ from $p(\mathbf{u})$
\For{$b=1,2,\ldots$}
\State Probe $\mathcal{S}_b$ and receive $\mathbf{y}_b$ by (\ref{eq:pilot})
\State Weight $\{\mathbf{u}^{(p)}\}$ by (\ref{eq:loglik}) and resample
\State Apply $T$ moves with (\ref{eq:mh}) to every particle
\State Select $\hat n_b$ by (\ref{eq:select}) with (\ref{eq:portpost})
\EndFor
\end{algorithmic}
\end{algorithm}

Algorithm~\ref{alg:smc} summarizes the proposed scheme. The likelihood (\ref{eq:loglik}) involves only $\hat L\times\hat L$ matrices, and the cost per block grows linearly with the numbers of stored blocks, moves, particles, and probes, and cubically with the number of model paths.

\subsection{Angle Information and Learning Speed}
The angles are a sufficient statistic of the past blocks for the current channel. Consequently, the selection by (\ref{eq:marcum}) at the true angles lower-bounds the outage probability of any scheme that relies only on the same probes. The speed at which the angle memory approaches the bound is set by the angle information of one block, which the following proposition characterizes for a single path.

\begin{proposition}\label{prop:fim}
For $L=1$, the Fisher information of the directional cosine $u$ carried by block $b$ is given by
\begin{equation}\label{eq:fim}
J_b(u)=\frac{8\pi^2k^2\,\varsigma_b^2}{\sigma^2(\sigma^2+k)},
\end{equation}
where $\varsigma_b^2=\frac{1}{k}\sum_{n\in\mathcal{S}_b}x_n^2-\big(\frac{1}{k}\sum_{n\in\mathcal{S}_b}x_n\big)^2$ is the spread of the probed positions.
\end{proposition}
\begin{proof}
For a single path, the pilot vector (\ref{eq:pilot}) is zero-mean Gaussian with an angle-dependent covariance, and the Fisher information follows from \cite{kay1993estimation} as
\begin{equation}\label{eq:fimproof}
J_b(u)=\mathrm{tr}\big(\mathbf{C}^{-1}\dot{\mathbf{C}}\,\mathbf{C}^{-1}\dot{\mathbf{C}}\big),
\end{equation}
where $\mathbf{C}=\mathbf{a}_{\mathcal{S}}\mathbf{a}_{\mathcal{S}}^{H}+\sigma^2\mathbf{I}_k$ is the covariance of the pilot vector with $\mathbf{a}_{\mathcal{S}}=[\mathbf{a}(u)]_{\mathcal{S}_b}$. Moreover, $\dot{\mathbf{C}}=\dot{\mathbf{a}}_{\mathcal{S}}\mathbf{a}_{\mathcal{S}}^{H}+\mathbf{a}_{\mathcal{S}}\dot{\mathbf{a}}_{\mathcal{S}}^{H}$ is the derivative of $\mathbf{C}$ with respect to $u$, where the derivative $\dot{\mathbf{a}}_{\mathcal{S}}$ has the entries $j2\pi x_n e^{j2\pi x_n u}$. The trace in (\ref{eq:fimproof}) expands as
\begin{equation}\label{eq:fimexpand}
J_b(u)=2\,\mathrm{Re}\big\{(\mathbf{a}_{\mathcal{S}}^{H}\mathbf{C}^{-1}\dot{\mathbf{a}}_{\mathcal{S}})^2\big\}+2\,(\mathbf{a}_{\mathcal{S}}^{H}\mathbf{C}^{-1}\mathbf{a}_{\mathcal{S}})(\dot{\mathbf{a}}_{\mathcal{S}}^{H}\mathbf{C}^{-1}\dot{\mathbf{a}}_{\mathcal{S}}).
\end{equation}
As every entry of $\mathbf{a}_{\mathcal{S}}$ has unit modulus, the matrix inversion lemma gives $\mathbf{C}^{-1}=\sigma^{-2}\big(\mathbf{I}_k-\mathbf{a}_{\mathcal{S}}\mathbf{a}_{\mathcal{S}}^{H}/(\sigma^2+k)\big)$. Accordingly, the three quadratic forms in (\ref{eq:fimexpand}) are obtained as
\begin{subequations}\label{eq:fimforms}
\begin{align}
\mathbf{a}_{\mathcal{S}}^{H}\mathbf{C}^{-1}\mathbf{a}_{\mathcal{S}}&=\frac{k}{\sigma^2+k},\label{eq:fimforms_a}\\
\mathbf{a}_{\mathcal{S}}^{H}\mathbf{C}^{-1}\dot{\mathbf{a}}_{\mathcal{S}}&=\frac{j2\pi s_1}{\sigma^2+k},\label{eq:fimforms_b}\\
\dot{\mathbf{a}}_{\mathcal{S}}^{H}\mathbf{C}^{-1}\dot{\mathbf{a}}_{\mathcal{S}}&=\frac{4\pi^2}{\sigma^2}\Big(s_2-\frac{s_1^2}{\sigma^2+k}\Big),\label{eq:fimforms_c}
\end{align}
\end{subequations}
where $s_1=\sum_{n\in\mathcal{S}_b}x_n$ and $s_2=\sum_{n\in\mathcal{S}_b}x_n^2$. Substitution of (\ref{eq:fimforms}) into (\ref{eq:fimexpand}) gives $J_b(u)=8\pi^2(ks_2-s_1^2)/(\sigma^2(\sigma^2+k))$, which equals (\ref{eq:fim}) by $ks_2-s_1^2=k^2\varsigma_b^2$.
\end{proof}

\begin{remark}\label{rem:design}
The angle information of one block has three properties. 1) By (\ref{eq:fim}), one probe per block carries no angle information, given $\varsigma_b^2=0$ for $k=1$. For any number of paths, the same follows from (\ref{eq:pilot}), since one probe follows $\mathcal{CN}(0,1+\sigma^2)$ for all angles. 2) As (\ref{eq:fim}) shows, the information of two or more probes grows with the spread of the probes. The information grows linearly with the number of probes at high pilot SNR and quadratically at low pilot SNR, where it decays with the square of the noise power. 3) For a single path and two fixed probes, the likelihood depends on $u$ only through the phase difference of the two probes. The resulting ambiguity repeats when the directional cosine changes by the inverse of the probe distance, and the product in (\ref{eq:post}) over random probe sets resolves it.
\end{remark}

Given the angle, the pilot vectors of different blocks are independent, and the Fisher information adds up over blocks. It follows that any unbiased estimate $\hat u$ of $u$ from $b$ blocks satisfies the Cram\'er--Rao bound (CRB) given by
\begin{equation}\label{eq:crb}
\mathbb{E}\big[(\hat u-u)^2\big]\ge\Big(\sum_{i=1}^{b}J_i(u)\Big)^{-1},
\end{equation}
which decays inversely with the number of blocks and quantifies the gain of the angle memory. Therefore, the number of blocks for a given accuracy scales inversely with (\ref{eq:fim}).

For several paths, the angle memory is even necessary. One block provides $2k$ real observations for $L$ angles and $2L$ real gain components. As a result, memoryless estimators cannot identify the angles in general if $2k<3L$, which holds for four probes and three paths. The angle memory avoids this limit by combining the probes of many blocks for the same angles.

\section{Simulation Results}
Unless otherwise specified, the simulation parameters are listed in Table~\ref{tab:param}. In particular, a FAS of $32$ ports over $4$ wavelengths receives three paths, and the proposed scheme adopts six model paths, twice the true number. For pilot and data SNRs of $10$~dB, the target rate of $1$~bit/s/Hz yields an outage probability of about $0.1$ for a fixed port. All results are averaged over independent realizations, and the steady state refers to the last blocks of each realization. Four benchmarks with the same probes are considered. 1) \textbf{Best probed}: The probed port with the largest pilot power is selected. 2) \textbf{LMMSE}: The memoryless LMMSE scheme \cite{skouroumounis2023lmmse} applies (\ref{eq:optsel}) under the isotropic correlation $\sin(2\pi d)/(2\pi d)$ of two ports at distance $d$. 3) \textbf{Bayes}: Algorithm~\ref{alg:smc} runs with the history reset in every block. The resulting scheme approximates the best memoryless selection and serves as a proxy for learned and generative priors \cite{tang2025diffusion}. 4) \textbf{Known angles}: The lower bound with the true angles. Since every port has the same average power, a selection based on the average power over past blocks gains nothing from memory. Accordingly, the gain of the proposed scheme stems from the angle model.

\begin{table}[t]
\centering
\caption{Simulation Parameters}
\label{tab:param}
\begin{tabular}{@{}ll@{}}
\toprule
Parameter & Value\\
\midrule
Ports, aperture, paths $L$, model paths $\hat L$ & $32$, $4\lambda$, $3$, $6$\\
Pilot SNR, data SNR, target rate & $10$~dB, $10$~dB, $1$~bit/s/Hz\\
Particles, moves, step sizes, prior draw & $2048$, $10$, $0.03$/$0.006$, $0.15$\\
Realizations, blocks & $4000$, $50$ ($3000$, $30$ in Fig.~\ref{fig:robust}(b))\\
\bottomrule
\end{tabular}
\end{table}

\begin{figure}[t]
\centering
\subfloat[Root-mean-square error of the angle estimate for a single path.]{\includegraphics[width=\columnwidth]{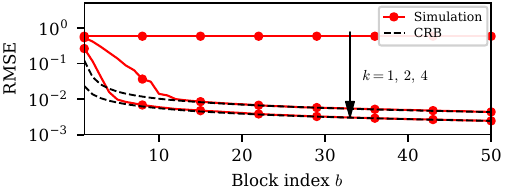}\label{fig:crb}}\\[1mm]
\subfloat[Outage probability for three paths and $k=4$.]{\includegraphics[width=\columnwidth]{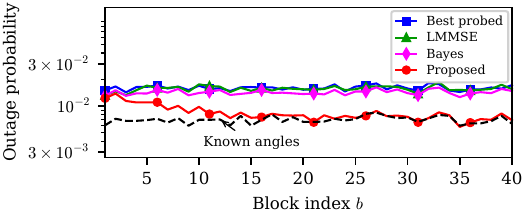}\label{fig:block}}
\caption{Angle learning and its effect over blocks.}
\label{fig:main}
\end{figure}

Fig.~\ref{fig:main}(a) verifies Proposition~\ref{prop:fim} for a single path, where the posterior (\ref{eq:post}) is evaluated on a fine grid over $2000$ realizations and its mean serves as the angle estimate. For one probe per block, the root-mean-square error (RMSE) stays at the level of the prior, which confirms Remark~\ref{rem:design}. For two and four probes, the RMSE first exceeds the CRB (\ref{eq:crb}), since the posterior mean is biased while the ambiguity of the phase differences is unresolved. After the ambiguity is resolved, the RMSE merges with the bound within about fifteen blocks for two probes and ten for four. These results confirm that the angle information accumulates at the rate predicted by (\ref{eq:fim}).

Fig.~\ref{fig:main}(b) depicts the outage probability versus the block index, averaged over four independent runs. In the first block, both schemes start from the same prior without history, and the proposed scheme nearly coincides with the memoryless Bayes scheme. As blocks accumulate, the outage probability of the proposed scheme decreases and approaches the lower bound after about twenty blocks. In contrast, all memoryless schemes stay flat, as every block starts again from the same prior. Among them, the memoryless Bayes scheme improves only slightly on the best probed port, since four probes cannot resolve three paths within one block. As a result, the past probes reveal the angles and enable every later block to predict the unprobed ports.

\begin{figure}[t]
\centering
\subfloat[Versus the number of probes in blocks $31$--$50$.]{\includegraphics[width=\columnwidth]{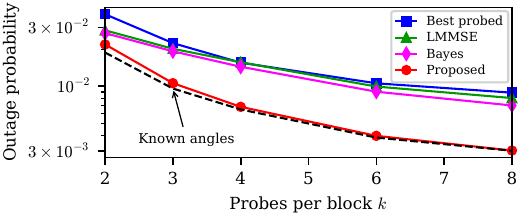}\label{fig:k}}\\[1mm]
\subfloat[Versus the angular spread of clustered paths for $k=4$ in blocks $16$--$30$.]{\includegraphics[width=\columnwidth]{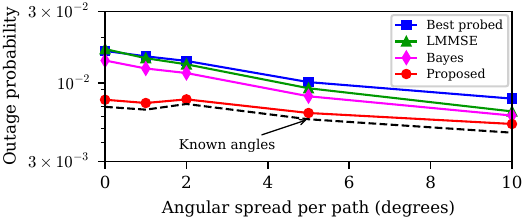}\label{fig:spread}}
\caption{Steady-state outage probability.}
\label{fig:robust}
\end{figure}

Fig.~\ref{fig:robust}(a) plots the steady-state outage probability versus the number of probes. For two probes, a visible gap to the lower bound remains, since each block carries little angle information by (\ref{eq:fim}). As probes are added, the gap shrinks steadily and nearly vanishes for eight probes. Compared with the memoryless schemes, the proposed scheme roughly halves the outage probability of the best one for three or more probes. In particular, the proposed scheme with four probes performs comparably to the memoryless schemes with eight probes. Consequently, a target outage probability is met with fewer probes per block and a shorter training phase.

Fig.~\ref{fig:robust}(b) evaluates the robustness against clustered scattering, where each of the three paths is replaced by a cluster of $10$ rays with independent gains and a Gaussian angular profile. Accordingly, the lower bound uses the angles of all $30$ rays. Although six model paths only approximate $30$ rays, the proposed scheme retains a clear gain for all angular spreads, and the gap to the lower bound stays below about $15\%$. Meanwhile, the memoryless schemes improve with the spread, as richer scattering makes the probed ports more independent. The gain of the proposed scheme narrows accordingly but does not vanish.

\begin{table}[t]
\centering
\caption{Steady-State Outage Probability ($\times 10^{-3}$)}
\label{tab:robust}
\begin{tabular}{@{}lccc@{}}
\toprule
Condition & Best memoryless & Proposed & Known angles\\
\midrule
Default & $14.1$ & $7.7$ & $6.9$\\
Pilot SNR $0$~dB & $56.6$ & $48.7$ & $40.4$\\
Pilot SNR $20$~dB & $4.5$ & $2.0$ & $2.0$\\
Paths $L=2$ & $28.5$ & $16.8$ & $16.3$\\
Paths $L=5$ & $7.5$ & $5.0$ & $4.9$\\
Drift $0.002$ & $14.7$ & $7.7$ & $6.9$\\
Drift $0.005$ & $14.1$ & $7.9$ & $7.3$\\
Drift $0.01$ & $14.7$ & $8.6$ & $7.6$\\
\bottomrule
\end{tabular}
\end{table}

Table~\ref{tab:robust} lists the steady-state outage probability when one condition of Table~\ref{tab:param} is changed, for four probes and the realizations and blocks of Fig.~\ref{fig:robust}(b). For drifting angles, each directional cosine and the particles receive independent Gaussian increments of the given standard deviation in every block. The past likelihoods are re-evaluated at the moved angles, an approximation that is exact only for static angles. At a pilot SNR of $0$~dB, the gap to the lower bound widens to about $20\%$, as (\ref{eq:fim}) decreases with the noise power. Even so, the proposed scheme lies about $14\%$ below the best memoryless scheme, and this margin grows to about $55\%$ at $20$~dB. For two and five true paths, the six model paths keep the gap to the lower bound within about $3\%$. Under drift, the outage probability barely changes up to $0.005$ per block and rises by about $12\%$ for $0.01$, which is about $0.6^\circ$ near broadside. Nevertheless, it stays about $40\%$ below the best memoryless scheme. Consequently, the proposed scheme tolerates an unknown number of paths and requires the geometry to persist only over tens of blocks.

\section{Conclusion}
In this paper, we have proposed a two-timescale channel estimation and port selection scheme for FAS with limited port observations. Specifically, a sequential Monte Carlo filter accumulates the posterior of the slowly varying path angles over blocks, while the fast-fading gains are integrated out in closed form. Based on the Fisher information, we have shown that the angle is learned only from the phase differences between spread-out probes. The learning speed is set by the number of probes and the pilot SNR. In addition, a count of the unknowns has revealed that the few probes of one block are in general insufficient for several paths. Simulation results have demonstrated that the proposed scheme approaches the lower bound of known angles and, for four probes per block, matches memoryless schemes with twice as many probes. Furthermore, the gain persists under an unknown number of paths, clustered scattering, and drifting angles. These results suggest that FAS training should exploit the geometry shared across blocks rather than treat each block independently. Future work will consider unequal path powers, a dynamic model of the angles, planar FAS, and probe set design.

\bibliographystyle{IEEEtran}
\bibliography{refs}
\end{document}